\documentclass[aps,physrev, reprint, groupedaddress]{revtex4-2}
\usepackage{amsmath}
\usepackage{amssymb}
\usepackage{graphicx}
\usepackage{dcolumn}
\usepackage{bm}
\usepackage{float}
\usepackage[bookmarks=false]{hyperref}
\usepackage{orcidlink}   
\newcommand{\Mn}{M_{n1}}
\newcommand{\chim}{\chi_{\max}}
\newcommand{\chinf}{\chi_{\infty}}
\newcommand{\Rinf}{R_{\infty}}
\newcommand{\tauz}{\tau_0}

\begin{document}
\title{Relativistic Oblique Shocks at Finite Temperature: Detachment Angle, Shock Polars, and the Turning Parameter}
\author{Rushikesh~Ashok~Sonkusale \orcidlink{0009-0004-3227-3748}}
\email{msc2503121010@iiti.ac.in}
\affiliation{Department of Astronomy, Astrophysics and Space Engineering,
                 Indian Institute of Technology Indore,
                 Simrol, Khandwa Road, Indore~453\,552, India}
                 
\author{Anshuman Verma \orcidlink{0000-0003-1103-0742}}
\email{ansshumanverma@gmail.com}
\affiliation{Indian Institute of Astrophysics,
                 II Block, Koramangala,
                 Bengaluru~560\,034, Karnataka, India}
                 
\author{Ritam Mallick \orcidlink{0000-0003-2943-6388}}
\email{mallick@iiserb.ac.in}
\affiliation{Department of Physics,
                 Indian Institute of Science Education and Research Bhopal,
                 Bhopal Bypass Road, Bhauri, Bhopal~462\,066, India}
\begin{abstract}
Oblique shocks are ubiquitous in high-energy astrophysical environments, yet a systematic analytical treatment of how finite upstream temperature influences the maximum deflection angle has been lacking. We address this problem by developing a unified thermodynamic framework based on a novel dimensionless quantity, the turning parameter, which encapsulates the equation of state, upstream Mach number, and thermal state of the flow into a single variable. Starting from the relativistic Rankine–Hugoniot conditions and the Taub adiabat, we derive a compact turning relation and a first-order perturbative expansion in the upstream thermal parameter. We show that any finite upstream temperature monotonically suppresses the maximum deflection angle relative to the cold-fluid limit, implying that cold models systematically overestimate shock attachment. In the combined ultra-thermal and ultra-relativistic limit, the turning parameter saturates to a universal value, yielding an asymptotic detachment angle that depends only on the equation of state. Numerical shock-polar calculations validate the analytical results and reveal a non-monotonic dependence of the detachment angle on the Mach number at intermediate temperatures, arising from the competition between thermal pressure and bulk kinetic energy—a distinctly relativistic thermal effect absent in both the cold and ultra-hot limits. As an illustrative astrophysical application, we apply the framework to the Crab pulsar wind nebula, demonstrating how finite-temperature effects modify the termination-shock morphology and the observed torus geometry.
\end{abstract}
\maketitle
\section{Introduction}
\label{sec:intro}
Shock waves are one of the most frequent phenomena in an astrophysical environment \cite{Anile1989, Landau1987}. Starting from core-collapse supernova \cite{Chevalier1982, Kamijima2024, Petruk2026, Cristofari2025}, gamma ray bursts \cite{Blandford1974, Daigne1998, Murase2006, Tavani2025, Charlet2025}, phase transition in compact objects \cite{ritam_2006,ritam-igor_2014,prasad_2018,irfan_2019,ritam_2021,anshuman_2022,anshuman_2023} to heavy-ion collision \cite{Csernai1988, Cissoko1992}.
Among them, oblique shocks are of special importance, as they are among the most studied structures in high-energy astrophysical environments. They shape the blast waves generated by a core-collapse supernova \cite{Chevalier1982}. They also act as a collimating agent and re-accelerator in the relativistic jets produced in active galactic nuclei (AGN) and gamma-ray burst central engines \cite{Blandford1974, Daigne1998, Murase2006, Senno2016, Peretti2026, SaizPerez2025} and in phase transition from hadronic to quark matter in neutron stars \cite{ritam-stefan_2014,ritam_2019}. Unlike normal shocks, they introduce a deflection angle ($\chi$) in the downstream flow, determined jointly by shock angle ($\phi$) and the upstream thermodynamic state \cite{Drury1983, Landau1987, Courant1948}. The locus of all the possible deflections, often represented by a shock polar diagram, reaches its apex at the maximum deflection angle $(\chi_{\mathrm{max}})$. If the required deflection by the downstream flow is larger than $\chi_{\mathrm{max}}$, then there is no solution with an oblique shock; it stands off as a bow shock, entering into the physics of dissipation and particle acceleration \cite{Blandford1976, Drury1983, Decker1985, Decker2008, Huang2023, Lemoine2025}.
The cold limit of shock physics is well understood. In both the Newtonian and relativistic regimes, the full landscape of admissible shock configurations has been systematically mapped \cite{Taub1948}, and the resulting framework has served the community well for decades. However, real astrophysical plasmas are not cold. In the astrophysical systems, ranging from supernova remnant to relativistic jet working surfaces, the ratio $p/\rho c^2$ spans from $0.1$ to $10$ \cite{Anile1989, Rezzolla2013}, which is far from the cold fluid approximation. In this regime, thermal pressure is not a small correction one can safely neglect; it actively reshapes the enthalpy balance across the shock front \cite{Taub1948, Rezzolla2013}, and with it, the entire geometry of shock deflection. To handle this, we introduce a quantity, the thermal parameter $\alpha_1$, defined as the ratio of pressure to proper rest-mass density. This parameter absorbs all the thermodynamic complexity about the temperature profile.
A complementary overview on this problem was stated by \citet{Shi2020}, where they approached the shock deflection problem purely from the geometric perspective; their analysis showed that the shock intensity gradient vanishes along the locus, a result that is purely independent of the polytropic index $\Gamma$, hence stating it to be universal. Elegant as this is, the thermodynamic underpinning of the geometric condition was never established; it remained an observed regularity without a physical explanation rooted in the microphysics of the shock. The present paper closes that gap. We develop a unified analytic framework built around two carefully chosen quantities. The first is the thermal parameter $\alpha_1$. The second and, as it turns out, the more fundamental of the two, is what we call the \textit{turning parameter R}, defined as
\begin{equation}
    R \equiv \frac{h_2 / \tau}{h_1},
    \label{eq:R_def}
\end{equation}
where $h_1$ and $h_2$ are the specific enthalpies in the upstream and downstream states respectively, $\tau$ is the compression ratio across the shock. $R$ measures the specific enthalpy carried downstream relative to that of the upstream flow, thereby quantifying the thermodynamic work performed by the shock. It serves as a sufficient statistic for the shock: the effects of the equation of state, the upstream Mach number, and the upstream thermal state are all encapsulated within this single parameter. We demonstrate that $R$ alone determines the complete deflection geometry of a relativistic oblique shock. Viewed from this perspective, the thermodynamic origin of the Shi et al. \cite{Shi2020} locus becomes transparent, with the entire family of shock solutions emerging as a direct consequence of the enthalpy transformation encoded in $R$.
Starting from the first principles, we derive a sequence of results that together form a self-sufficient analytic picture of the oblique shocks at a given finite temperature. Starting with the Taub adiabat \cite{Taub1948}, the relativistic analogue of the classical Rankine-Hugoniot condition, it connects the enthalpy and pressure jumps across the shock, using which one derive a turning relation expressed entirely in terms of the turning parameter $R$. 
Moving beyond the cold limit, we derive the first-order expansion of $\alpha_1$ to quantify the influence of a finite upstream temperature on the detachment angle. We show that any finite upstream thermal content suppresses the maximum deflection angle, $\chi_{\mathrm{max}}$. At the opposite extreme, by considering the combined ultra-thermal and ultra-relativistic limit, we uncover a universal asymptotic detachment angle,
\begin{equation*}
    \chi_{\infty} = \arcsin\!\left(\frac{2 - \Gamma}{\Gamma}\right),
\end{equation*}
which emerges naturally from the saturation of the turning parameter $R$. Finally, our generalized thermal framework recovers the minimum-intensity locus, $\phi=\chi/2+\pi/4$, identified by Shi et al., demonstrating that it arises as a direct consequence of the underlying shock thermodynamics and thereby providing a clear physical interpretation of its origin.

The paper is organised as follows. In Sec. II, we present the formalism underlying our analysis and derive the key analytical results from first principles. In Sec. III, we present the numerical results and illustrate the applicability of our formalism through a case study of the Crab Pulsar Wind Nebula. Finally, in Sec. IV, we summarise our findings and present our conclusions.

\section{Formalism}
\label{sec:formalism}
\subsection{Conservation Laws}
\label{sec:conservation}
\begin{figure}
  \centering
  \includegraphics[width=\columnwidth]{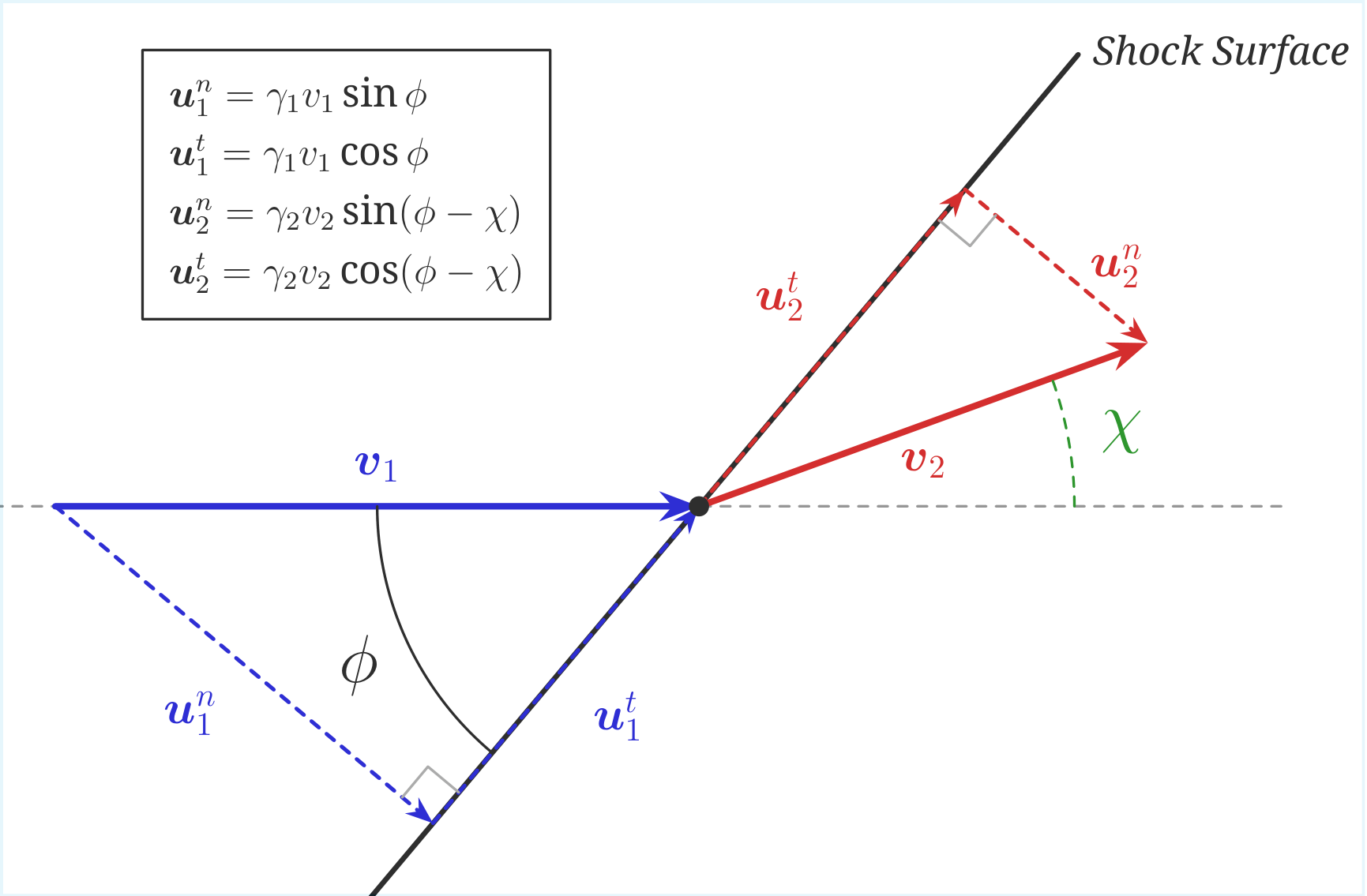}
  \caption{Shock geometry in the shock rest frame. The shock angle $\phi$ is measured from the shock surface, giving $u^n=\gamma v\sin\phi$ and $u^t=\gamma v\cos\phi$. The flow is deflected by $\chi$ on crossing the shock; the tangential component of momentum remains continuous, while the normal component is reduced by compression. Subscripts $1$ and $2$ label the upstream and downstream velocity decompositions, respectively, same as used in Eqs.~(\ref{eq:vel_components})--(\ref{eq:turning}) below.}
  \label{fig:geometry}
\end{figure}
Throughout this work, all calculations are performed in the shock rest frame. We adopt natural units with $c=1$ and use the Minkowski metric convention $\eta_{\mu\nu}=\mathrm{diag}(-1,+1,+1,+1)$. Subscripts 1 and 2 denote upstream and downstream quantities, respectively. We define the upstream thermal parameter as $\alpha_1 \equiv p_1/\rho_1$, the ratio of thermal pressure to rest-mass energy density, which serves as a convenient measure of the upstream temperature. Using the ideal-gas relations $p=nkT$ and $\rho=mn$, one obtains $\alpha=kT/m$ in natural units \cite{Rezzolla2013}. The upstream relativistic adiabatic sound speed is given by $c_{s1}^2=\Gamma p_1/(\rho_1 h_1)$ \cite{Rezzolla2013, Johnson1971}, and the corresponding normal Mach number is defined as $M_n=M_1\sin\phi$. A summary of the principal notation used throughout the paper is provided in Table~\ref{tab:notation}.
\begin{table}
\caption{Principal symbols.\label{tab:notation}}
\begin{ruledtabular}
\begin{tabular}{lll}
Symbol       & Meaning                                  & Definition           \\
$\rho_i$     & proper rest-mass density                 &                      \\
$p_i$        & pressure                                 &                      \\
$h_i$        & specific enthalpy                        & $1+ap_i/\rho_i$      \\
$a$          & EOS shorthand                            & $\Gamma/(\Gamma-1)$  \\
$\alpha_1$   & upstream thermal parameter               & $p_1/\rho_1$         \\
$\tau$       & compression ratio                        & $\rho_2/\rho_1$      \\
$\xi$        & pressure ratio                           & $p_2/p_1$            \\
$j$          & invariant mass flux                      & $\rho_1 u^n_1$       \\
$R$          & turning parameter                        & $(h_2/\tau)/h_1$     \\
$\phi$       & shock angle (from shock surface) &                      \\
$\chi$       & flow deflection angle                    &                      \\
\end{tabular}
\end{ruledtabular}
\end{table}
Assuming a polytropic equation of state with a constant adiabatic index $\Gamma$, the specific enthalpy is given by
\begin{equation}
h_i = 1 + a\,\frac{p_i}{\rho_i}, \qquad
a \equiv \frac{\Gamma}{\Gamma-1},
\label{eq:eos}
\end{equation}
such that the upstream and downstream specific enthalpies become
$h_1=1+a\alpha_1$ and $h_2=1+a\xi\alpha_1/\tau$
where $\xi\equiv p_2/p_1$ denotes the pressure ratio \cite{Taub1948, Rezzolla2013, Johnson1971}.
Integrating baryon-number conservation $\partial_\mu(\rho u^\mu)=0$ across the shock yields the invariant mass flux
\begin{equation}
  \rho_1 u^n_1 = \rho_2 u^n_2 \equiv j, \qquad u^n_i = j/\rho_i.
  \label{eq:baryon}
\end{equation}
For a perfect relativistic fluid,
$T^{\mu\nu}=\rho h\,u^\mu u^\nu+p\,\eta^{\mu\nu}$, the energy-momentum conservation gives
\begin{align}
j^2\left(\frac{h_1}{\rho_1}-\frac{h_2}{\rho_2}\right)&=p_2-p_1,\label{eq:normalmom}\\
h_1u_1^t&=h_2u_2^t,\label{eq:tangmom}\\
h_1\gamma_1&=h_2\gamma_2.\label{eq:energy}
\end{align}
Eliminating the kinematic invariants using the four-velocity normalization leads to the Taub adiabat,
\begin{equation}
h_2^2-h_1^2
=\frac{p_2-p_1}{\rho_1}
\left(h_1+\frac{h_2}{\tau}\right),
\label{eq:taub}
\end{equation} the relativistic analogue of the Hugoniot relation, expressing the balance between enthalpy generation and compressional work.
Substituting Eq.~(\ref{eq:eos}) into Eq.~(\ref{eq:taub}) gives the master implicit equation,
\begin{equation}
\begin{split}
\left(1+\frac{a\xi\alpha_1}{\tau}\right)^{2}-(1+a\alpha_1)^2
&= \alpha_1(\xi-1)\Bigl[(1+a\alpha_1) \\
&\quad + \frac{1}{\tau}\left(1+\frac{a\xi\alpha_1}{\tau}\right)\Bigr],
\end{split}
\label{eq:master}
\end{equation}
which is solved numerically for $\tau(\xi,\alpha_1,\Gamma)$. In the cold limit ($\alpha_1\rightarrow0$), the master equation recovers the Rankine-Hugoniot compression ratio,
\begin{equation}
\tau_0=
\frac{(\Gamma+1)M_n^2}
{(\Gamma-1)M_n^2+2},
\label{eq:tau0}
\end{equation}
with $R_0=1/\tau_0$, thereby reproducing the standard relativistic oblique-shock relation of Landau-Lifshitz and Shi et al.~\cite{Shi2020}. This serves as the exact boundary condition for all subsequent results.
Fig.~\ref{fig:geometry} shows the geometry of an oblique shock. The shock angle $\phi$ is the upstream flow angle measured from the shock surface.
For an oblique shock, the upstream four-velocity components are
\begin{equation}
u^n_1 = \gamma_1 v_1\sin\phi, \qquad u^t_1 = \gamma_1 v_1\cos\phi.
\label{eq:vel_components}
\end{equation}
Combining Eqs.~(\ref{eq:baryon}) and (\ref{eq:tangmom}) yields the turning relation,
\begin{equation}
\sin(2\phi-\chi) = \frac{1+R}{1-R}\sin\chi, \quad
R \equiv \frac{h_2/\tau}{h_1}.
\label{eq:turning}
\end{equation}
where $R$ compares the downstream specific enthalpy transported per unit compressed mass with the upstream enthalpy supply. The equation of state, Mach number, compression, and upstream thermal state enter the shock geometry only through this single parameter, making $R$ the fundamental thermodynamic variable governing the shock polar. 
Since $(1+R)/(1-R)$ increases monotonically with $R$, a larger $ R$ requires a larger value of $\sin(2\phi-\chi)$ for a given $\sin\chi$, thereby compressing the shock polar and reducing the maximum attainable deflection angle. Physically, $R$ compares the downstream-specific enthalpy transported per unit mass compressed with the upstream enthalpy supplied, quantifying the thermodynamic work performed by the shock. Small values of $R$ correspond to strong compression with relatively little downstream heating, permitting larger flow deflections, whereas larger values of $ R$ indicate enhanced enthalpy generation that suppresses the deflection angle.

Defining $F(\phi,\chi) \equiv \sin(2\phi - \chi) - [(1+R)/(1-R)] \sin \chi$ the minimum admissible shock angle is $\phi_{\min}=\arcsin(1/M_1)$, which follows from the requirement that the normal Mach number satisfy $M_n=M_1\sin\phi>1$ \cite{Landau1987, Courant1948}. As $\phi$ varies from $\phi_{\min}$ to $\pi/2$, the solution set of $F=0$ traces the shock polar \cite{Courant1948, Landau1987}. The polar begins and ends at $\chi=0$: at $\phi=\phi_{\min}$ the shock degenerates into a Mach wave, while at normal incidence ($\phi=\pi/2$) symmetry again requires zero flow deflection.
The apex of the shock polar corresponds to the coalescence of the weak- and strong-shock branches and therefore defines the maximum attainable deflection angle, $\chi_{\mathrm{max}}$ \cite{Courant1948, Landau1987}. Beyond this point, no attached oblique-shock solution exists. Consequently, if the half-angle of an obstacle exceeds $\chi_{\mathrm{max}}$, the shock detaches from the surface and forms a standing bow shock, a configuration central to astrophysical dissipation and particle acceleration \cite{Blandford1976, Drury1983, Decker1985, Decker2008}.
The attached shock limit is determined from $F=0$ and $\frac{\partial F}{\partial\phi} = 0$, which gives
\begin{equation}
  2\cos(2\phi-\chi) = \frac{2}{(1-R)^2}\frac{\partial R}{\partial\phi}\sin\chi.
  \label{eq:stationary}
\end{equation}
Together with Eq.~(\ref{eq:turning}), this determines the detachment point $(\phi_{\rm det},\chi_{\max})$.
In the strong-shock limit ($M_n\to\infty$), Eq.~(\ref{eq:tau0}) approaches the familiar compression ceiling $\tau_0 \to (\Gamma+1)/(\Gamma-1)$, 
giving $\tau_0=4$ for $\Gamma=5/3$ and $\tau_0=7$ for $\Gamma=4/3$ \cite{Kennel1984, Kirk2009}. Thus, a cold polytropic gas admits a finite maximum compression ratio, a restriction that is lifted once finite upstream thermal effects ($\alpha_1>0$) are included (see Sec.~\ref{app:ultra}).
The corresponding cold-limit turning parameter, $R_0=1/\tau_0$, satisfies
\begin{equation}
\frac{1+R_0}{1-R_0}
=\frac{\Gamma M_n^2+1}{M_n^2-1},
\end{equation}
which, upon substitution into Eq.~(\ref{eq:turning}) with $R=R_0$, recovers the classical Landau-Lifshitz oblique-shock relation,
\begin{equation}
M_n^2=1+\frac{(\Gamma+1)\sin\chi}
{\sin(2\phi-\chi)-\Gamma\sin\chi}
\label{eq:ll_mach}
\end{equation}
in agreement with the results of \citet{Shi2020} (see also Appendix~\ref{app:ll}).

To quantify finite-temperature effects, we expand
\begin{align}
\tau &= \tauz + \alpha_1\tau_1 + O(\alpha_1^2), \notag \\
R &= R_0 + \alpha_1 R_1 + O(\alpha_1^2), \quad R_0 = 1/\tauz, \notag \\
\chim &= \chi_0 + \alpha_1\chi_1 + O(\alpha_1^2).
\label{perturb}
\end{align}
Hence, any finite upstream thermal content increases the turning parameter and suppresses the maximum attainable deflection angle. Physically, thermal energy contributes to downstream enthalpy rather than compression, reducing the momentum available for flow turning.
Matching $O(\alpha_1)$ terms of the master equation (Appendix~\ref{app:perturb}) yields
\begin{equation}
\tau_1=
\frac{a\tau_0^2(\xi-1+a)+a\xi\bigl[(\xi-1)-a\xi\bigr]}
{(\xi-1)-2a\xi},
\label{eq:tau1}
\end{equation}
where $\xi=\xi_{\rm cold}\equiv[2\Gamma\Mn^2-(\Gamma-1)]/(\Gamma+1)$.
The corresponding first-order correction to the turning parameter is
\begin{equation}
R_1=\frac{a\xi-\tau_1}{\tau_0^2} -\frac{a}{\tau_0}.
\label{eq:R1}
\end{equation}
As shown in Appendix~\ref{app:perturb}, $R_1 > 0$ for all physically admissible strong shocks. Consequently, any finite upstream thermal content increases the turning parameter above its cold-limit value, $R=1/\tau_0$. Physically, the downstream enthalpy increases more rapidly than the compression does, thereby increasing the thermodynamic work performed by the shock and increasing $R$. Thus, the cold-limit models overestimate the maximum deflection angle $\chim$ and thereby overestimate the threshold for shock detachment in warm plasmas.
In the combined ultra-thermal ($\alpha_1\rightarrow\infty$) and ultra-relativistic ($M_n\rightarrow\infty$) limit, the Taub adiabat gives $\xi \sim (\Gamma-1)\tau^2$
so that the turning parameter saturates at
\begin{equation}
R_\infty=\Gamma-1.
\label{eq:Rinf}
\end{equation}
Since $R$ becomes independent of the shock angle, $\partial R/\partial\phi=0$, and the stationarity condition immediately yields
\begin{equation}
\phi=\frac{\chi}{2}+\frac{\pi}{4},
\label{eq:minlocus}
\end{equation}
the minimum-intensity locus previously identified by Shi et al.~\cite{Shi2020}. Substituting this into the turning relation gives the universal asymptotic detachment angle,
\begin{equation}
\chi_\infty=\arcsin\left(\frac{2-\Gamma}{\Gamma}\right),
\label{eq:chiinf}
\end{equation}
which depends solely on the equation of state. Thus, the Shi et al. \cite{Shi2020} locus emerges naturally as the geometric consequence of the saturation of the thermodynamic turning parameter $R$, providing a transparent physical interpretation of its origin.
\section{Results}
\begin{figure*}
  \centering
  \includegraphics[width=\textwidth]{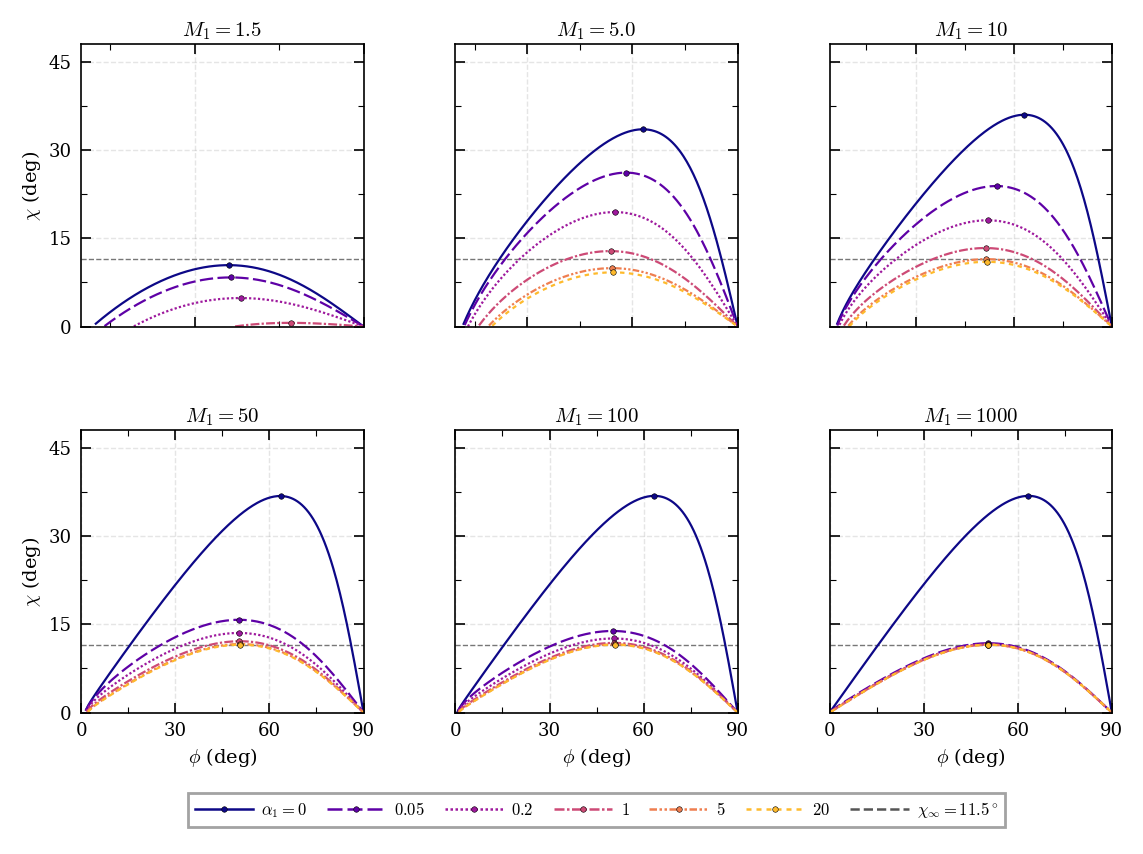}
  \caption{Shock-polar evolution with upstream Mach number ($\Gamma=5/3$). Each of the six panels shows the complete shock polar, deflection angle $\chi$ against shock angle $\phi$, for a fixed upstream Mach number $M_1=1.5$, $5.0$, $10$, $50$, $100$, and $1000$ (left to right, top to bottom), with $\phi$ ranging from $\phi_{\min}=\arcsin(1/M_1)$ to $\pi/2$. Within each panel, the six curves correspond to upstream thermal parameters $\alpha_1=0$ (cold limit), $0.05$, $0.2$, $1$, $5$, and $20$, distinguished by the line styles given in the legend; filled circles mark the detachment point $(\phi_{\rm det},\chim)$ on each curve. The horizontal dashed line in every panel marks the universal asymptotic detachment angle $\chinf=11.54^\circ$ from Eq.~(\ref{eq:chiinf}). Reading across the panels at fixed $\alpha_1$ shows the polar growing and $\chim$ increasing with $M_1$; reading down each panel at fixed $M_1$ shows the polar shrinking monotonically as $\alpha_1$ increases. At the highest Mach numbers the cold polar keeps growing without bound, while the finite-$\alpha_1$ polars flatten and converge toward, or in some cases fall below, the horizontal $\chinf$ line, depending on whether the combined ultra-thermal, ultra-relativistic limit of Appendix~\ref{app:ultra} has effectively been reached.}
  \label{fig:polargrid}
\end{figure*}
\label{sec:numerics}

\begin{figure}
  \centering
  \includegraphics[width=0.85\columnwidth]{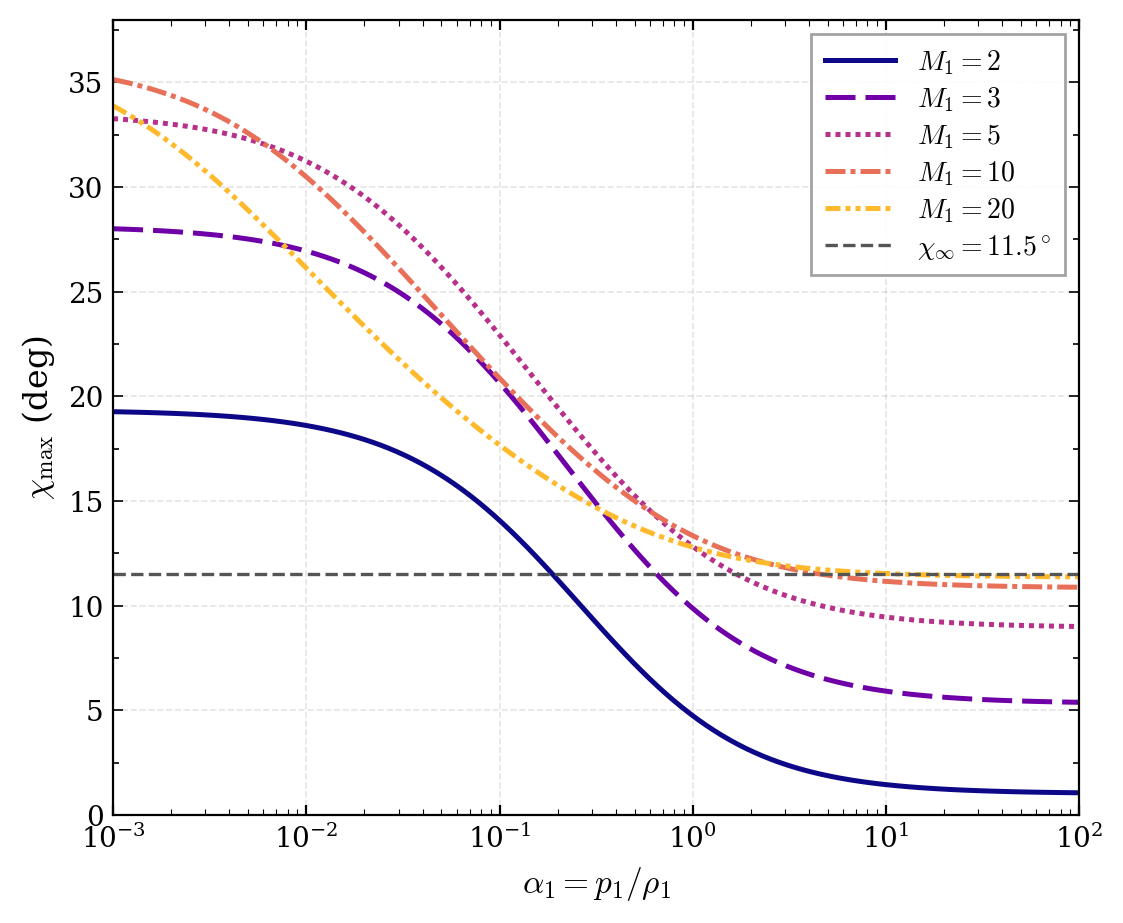}
  \caption{Maximum deflection angle $\chim$ vs.\ $\alpha_1$ for five Mach
    numbers ($\Gamma=5/3$).  The cold plateau is followed by monotonic
    descent. Larger-$M_1$ curves approach $\chinf=11.54^\circ$ (horizontal dashed)
    from above; lower-$M_1$ curves descend below $\chinf$ at large $\alpha_1$
    (see Sec.~\ref{app:ultra}). The near-flat plateau at small $\alpha_1$ reflects the
    quadratic onset of the thermal correction, $\chim(\alpha_1)\simeq\chi_0+\alpha_1\chi_1$,
    before the perturbative expansion of Eq.~(\ref{perturb}) eventually breaks down.}
  \label{fig:chimax_alpha}
\end{figure}

Shock polars are constructed by varying the upstream flow angle from $\phi_{\min}=\arcsin(1/M_1)$ to normal incidence ($\pi/2$). For each choice of $(M_1,\alpha_1,\Gamma)$, we compute $M_n=M_1\sin\phi$ and solve the coupled shock jump conditions. The master equation Eq.~(\ref{eq:master}) together with the pressure-ratio relation Eq.~(\ref{eq:B7}) is solved using Brent's method to obtain the compression ratio $\tau$ and pressure ratio $\xi$. The corresponding enthalpies determine the turning parameter,
$R=\frac{h_2/\tau}{h_1}$, which is substituted into Eq.~(\ref{eq:turning}) to obtain the deflection angle $\chi$. Repeating this procedure over the full range of $\phi$ traces the shock polar, while the detachment angle is identified as $\chi_{\max}=\max[\chi(\phi)]$.
\begin{figure}
  \centering
  \includegraphics[width=\columnwidth]{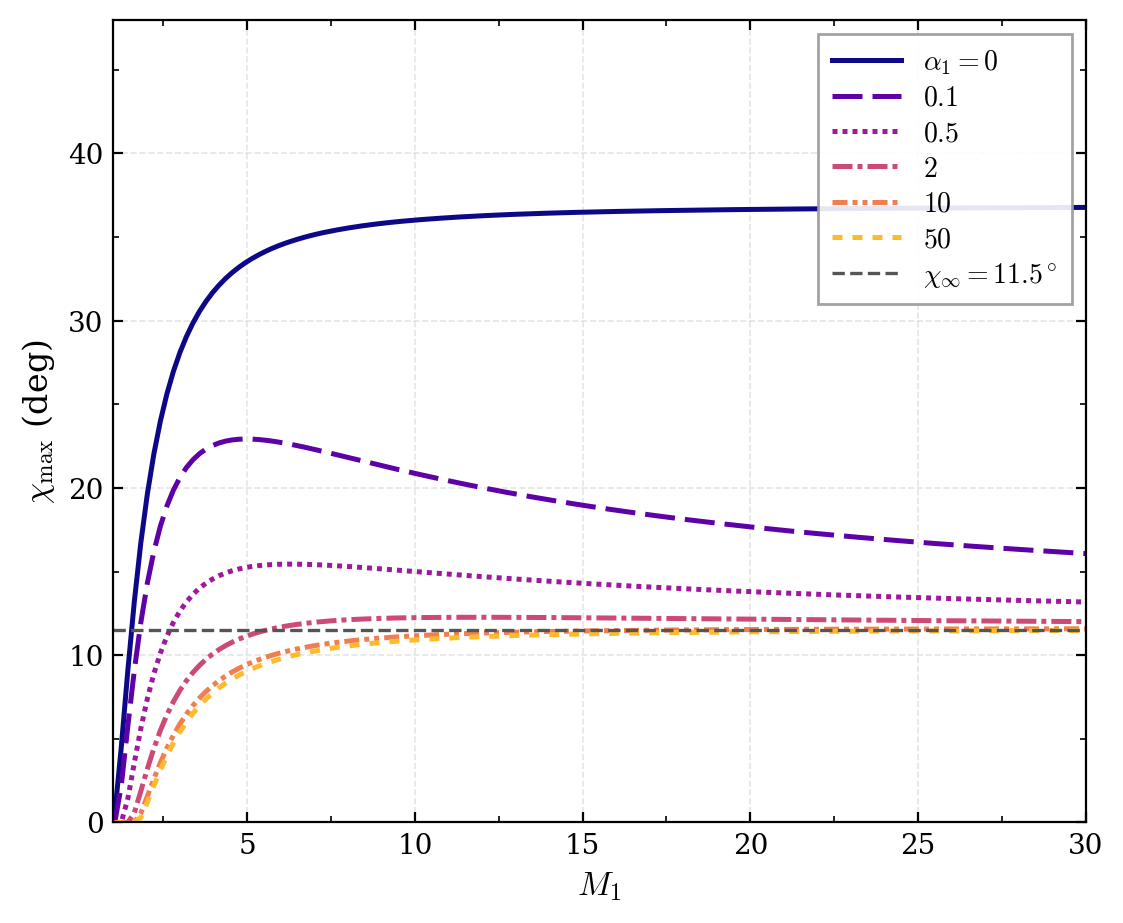}
  \caption{$\chim$ vs.\ $M_1$ for six values of $\alpha_1$ ($\Gamma=5/3$).
    The cold limit rises monotonically; all finite-temperature cases exhibit a
    non-monotonic rise-peak-fall.  Curves for $\alpha_1=2$ and $10$ converge
    to $\chinf$ (horizontal dashed) at large $M_1$.  This is a purely relativistic
    thermal effect with no Newtonian analogue. The location of the peak shifts to
    progressively lower $M_1$ as $\alpha_1$ increases, tracing out the competition
    between bulk kinetic energy and thermal pressure.}
  \label{fig:chimax_mach}
\end{figure}
\begin{figure}
  \centering
  \includegraphics[width=\columnwidth]{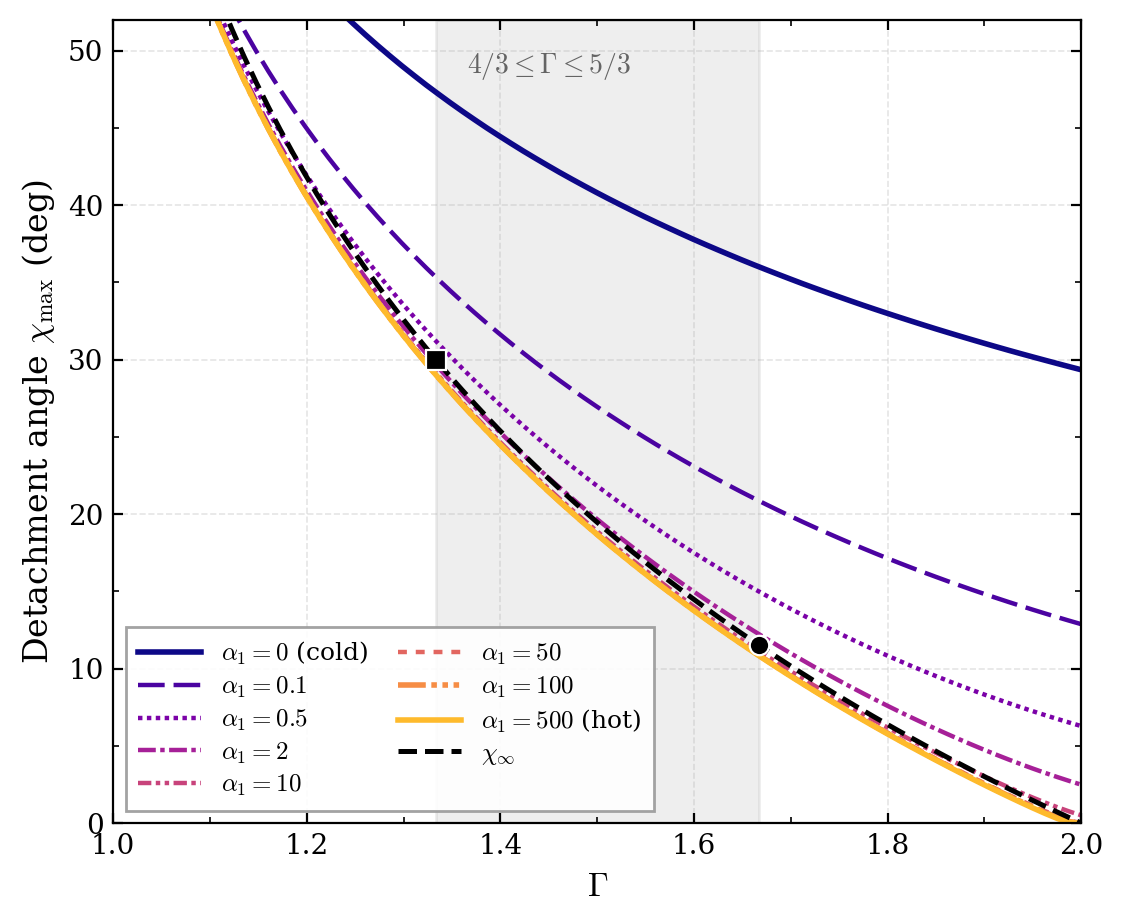}
  \caption{Detachment angle $\chim$ as a function of the adiabatic index $\Gamma$, at fixed upstream Mach number $M_1=10$, for eight upstream thermal parameters $\alpha_1=0$ (cold), $0.1$, $0.5$, $2$, $10$, $50$, $100$, and $500$ (hottest), distinguished by the line styles in the legend. The analytic ultra-thermal prediction $\chinf=\arcsin[(2-\Gamma)/\Gamma]$ of Eq.~(\ref{eq:chiinf}) (black dashed curve) agrees with the numerical $\alpha_1=500$ curve to plotting accuracy across the full range of $\Gamma$ shown, confirming that $\alpha_1=500$ is already deep enough into the ultra-thermal regime at $M_1=10$ to realize the asymptotic result. The grey band marks the astrophysically relevant range $4/3\leq\Gamma\leq5/3$, spanning radiation-dominated to non-relativistic monatomic gases, with black markers giving the exact detachment angle at the two boundary values of $\Gamma$.}
  \label{fig:eos}
\end{figure}

\begin{figure*}
  \centering
  \includegraphics[width=\textwidth]{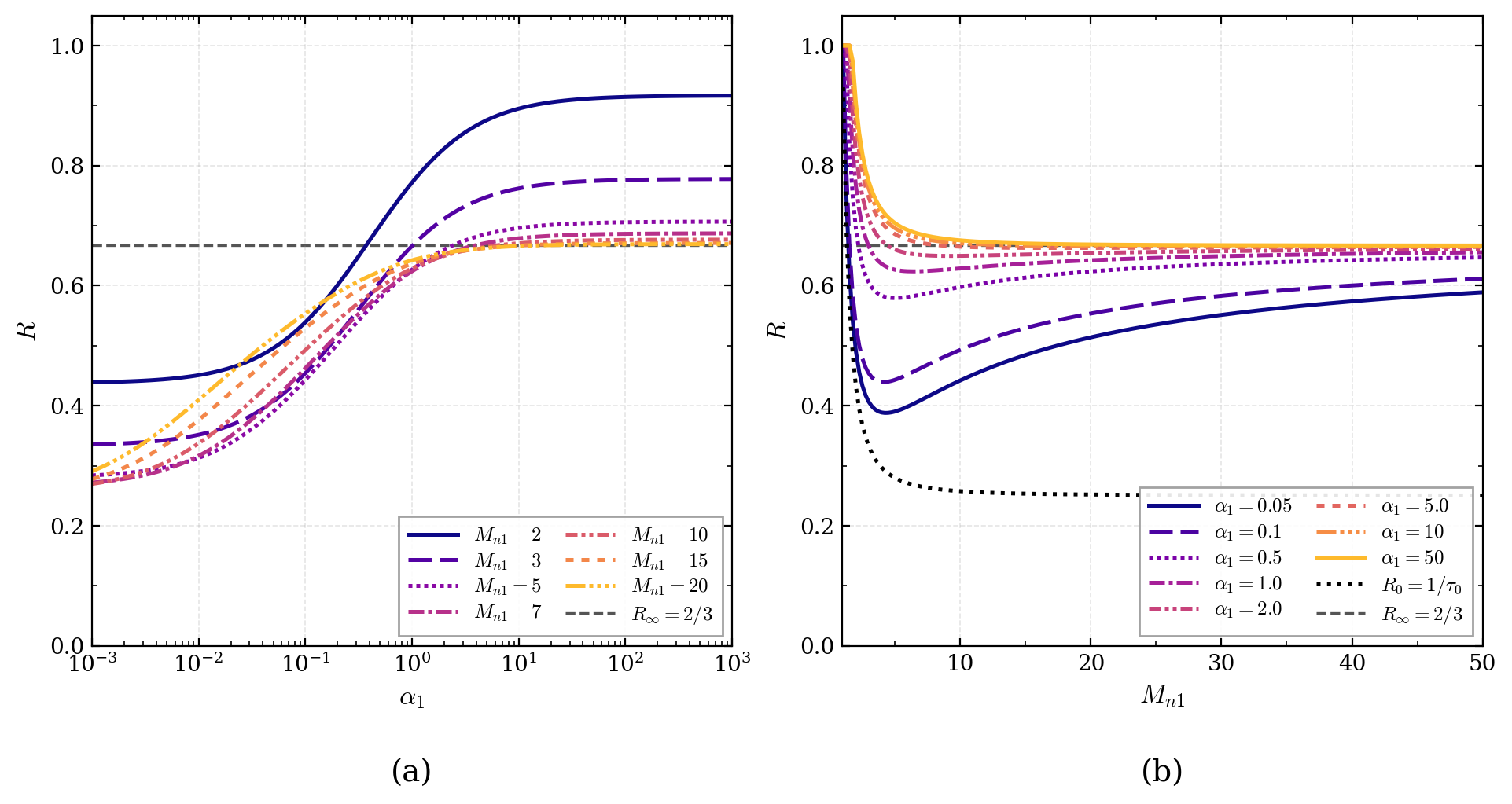}
  \caption{Convergence of the turning parameter $R$ to its ultra-thermal saturation $\Rinf=\Gamma-1=2/3$ (dashed, $\Gamma=5/3$). Left (a): $R$ vs upstream thermal parameter $\alpha_1$ (log scale) for $\Mn=2,3,5,7,10,15,20$; all curves rise from cold-limit to $\Rinf$ as $\alpha_1\to\infty$, except the $\Mn=2$ curve overshoots at intermediate $\alpha_1$ because the $\xi\to\infty$ approximation in Eq.~(\ref{eq:G2}) is inaccurate at low Mach. Right (b): $R$ vs $\Mn$ for $\alpha_1=0.05,0.1,0.5,1,2,5,10,50$; hot curves stay flat at $\Rinf$ over the entire range, while the cold-limit curve $R_0=1/\tauz$ (dotted) rises from below and saturates at $1/4$ as $\Mn\to\infty$ (Table~\ref{tab:summary}). Once $R=\Rinf$, the turning relation Eq.~(\ref{eq:turning}) is fully determined and $\chim$ is forced to its universal value $\chinf$, independent of other shock parameters.}
  \label{fig:R_conv}
\end{figure*}

Figure~\ref{fig:polargrid} shows shock polars for six Mach numbers ($1.5\le M_1\le5000$) and six upstream thermal parameters ($\alpha_1=0$, 0.05, 0.2, 1, 5, and 20). Two systematic trends emerge. First, for fixed $\alpha_1$, increasing the Mach number enlarges the shock polar and increases the attainable deflection angle. Second, for fixed $M_1$, increasing $\alpha_1$ monotonically shrinks the shock polar, in agreement with the perturbative prediction $\chi_1<0$. As both $M_1$ and $\alpha_1$ become large, the shock polars converge toward the universal asymptotic limit $\chi_\infty$, with the $M_1=5000$ curves being visually indistinguishable from the asymptotic solution.
The behavior is illustrated more clearly in the representative case $M_1=5$ (Fig.~\ref{fig:polar}, shown in Appendix~\ref{app:extra_figs}), where the shock-polar apex decreases monotonically with increasing $\alpha_1$, confirming the suppression of the maximum deflection by upstream thermal effects. Some curves lie below the asymptotic value $\chi_\infty$, reflecting that the ultra-thermal limit alone is insufficient; the asymptotic solution requires the simultaneous ultra-relativistic limit ($M_1\rightarrow\infty$).
The dependence of the detachment angle on the upstream thermal parameter is shown in Fig.~\ref{fig:chimax_alpha}. Each curve begins at the cold-limit value and decreases monotonically with increasing $\alpha_1$, approaching the asymptotic prediction $\chi_\infty$. The convergence is progressively faster for larger Mach numbers, consistent with the requirement that both $\alpha_1$ and $M_1$ be large. The complementary representation in Fig.~\ref{fig:chimax_mach} shows that the cold-limit deflection increases monotonically with Mach number, whereas finite-temperature flows exhibit a non-monotonic dependence before converging to the universal asymptote.
Figure~\ref{fig:compression} (Appendix~\ref{app:extra_figs}) demonstrates that finite upstream temperature removes the cold Rankine-Hugoniot compression ceiling, allowing the compression ratio to increase continuously. In realistic plasmas, however, additional microphysical processes—such as pair production and changes in the equation of state—ultimately limit the achievable compression \cite{Svensson1982, Lima2003}.

Finally, Fig.~\ref{fig:eos} confirms excellent agreement between the analytical cold-limit solution and the numerical calculations, while the high-temperature curves approach the universal prediction given by Eq.~(\ref{eq:chiinf}). The convergence of the turning parameter is shown explicitly in Fig.~\ref{fig:R_conv}: increasing either $\alpha_1$ or $M_1$ raises $R$ toward its asymptotic value $R_\infty=\Gamma-1$, while only their combined large-limit recovers the fully saturated thermodynamic regime predicted analytically in Sec.~\ref{app:ultra}.

Taken together, Figs.~\ref{fig:polargrid}--\ref{fig:R_conv} demonstrate that the turning parameter $R$ is not merely an auxiliary quantity but rather the fundamental thermodynamic variable that governs the entire family of shock solutions. The non-monotonic rise-peak-fall behaviour of $\chim(M_1)$ seen in Fig.~\ref{fig:chimax_mach} has a direct counterpart in the shock-polar grid of Fig.~\ref{fig:polargrid}. At fixed $\alpha_1$, the polar initially widens with increasing $M_1$ as bulk kinetic energy dominates the enthalpy balance; at higher Mach numbers, however, thermal enthalpy generation overtakes the compressional gain and pushes $R$ toward $\Rinf$ from below, causing the polar to contract. This crossover is most pronounced at intermediate $\alpha_1$ (Fig.~\ref{fig:chimax_alpha}), where neither the cold nor the ultra-thermal approximation is valid, and it disappears smoothly in both limits $\alpha_1\to0$ and $\alpha_1\to\infty$, in agreement with the perturbative and saturated limits derived in Sec.~\ref{sec:formalism}. The compression-ratio trends of Fig.~\ref{fig:compression} provide the thermodynamic underpinning of this crossover: lifting the cold ceiling $\tau_0\to(\Gamma+1)/(\Gamma-1)$ is precisely what allows $R$ to continue climbing toward $\Rinf$ at large $\Mn$, rather than saturating at the cold value $R_0$. Ultimately, it is this competition between rising compression and rising downstream enthalpy that determines the location of the non-monotonic peak in $\chim$.

\subsection{Application to the Crab pulsar wind nebula}
Pulsar wind nebulae (PWNe) provide an excellent observational laboratory for relativistic oblique shocks \cite{Gaensler2002, Buehler2016, Porth2017, Dinsmore2024, Olmi2024, Barkov2019, Toropina2019, Posselt2017, Gagnon2026, LHAASO2021, Cao2024, Li2024, Alispach2025, Spencer2025, Aharonian2024, Wei2025, Zuo2025, Martin2025, Chatterjee2025, Temim2024}. The Crab termination shock possesses a well-resolved geometry, while the upstream wind properties are constrained by the pulsar spin-down luminosity \cite{Kennel1984, Kirk2009, Buhler2014}. Since the Crab wind is strongly magnetized, the following discussion should be regarded as an illustrative thermodynamic application of our hydrodynamic framework rather than a complete RMHD model \cite{Porth2014, Camus2009, Lyutikov2018, Cerutti2014, Cerutti2020}.
\begin{figure}
  \centering
  \includegraphics[width=\columnwidth]{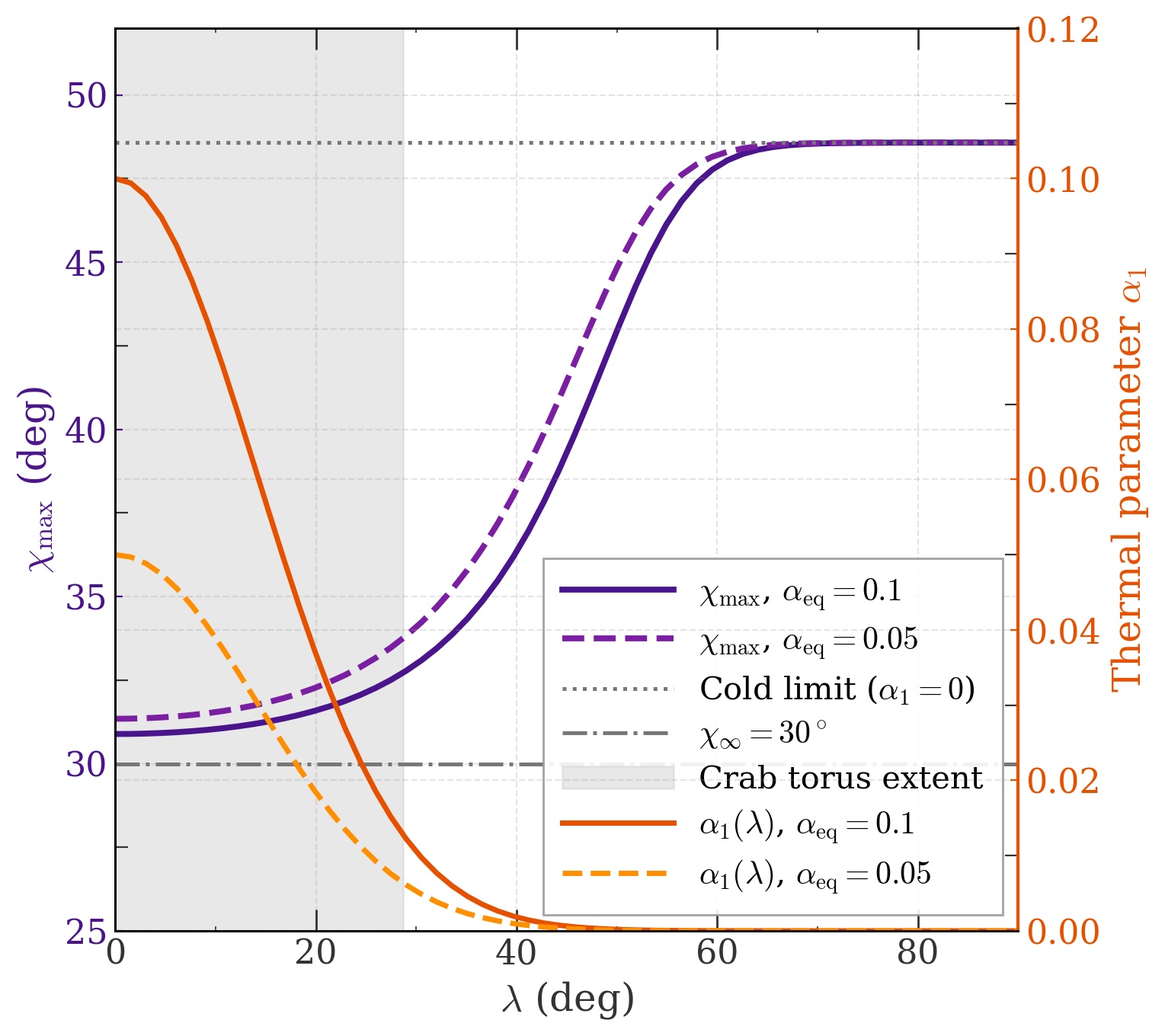}
  \caption{Latitude-dependent shock morphology. Left axis (purple): Maximum detachment angle \(\chi_{\text{max}}(\lambda)\) for two equatorial thermal parameters (\(\alpha_{\mathrm{eq}} = 0.1\), solid; and \(0.05\), dashed) at \(M_1=100\). Right axis (orange): Upstream thermal profile \(\alpha_1(\lambda)\) from Eq.~(\ref{eq:alpha_profile}) with corresponding line styles. The cold limit (\(\alpha_1=0\)) and universal asymptote (\(\chi_\infty=30^\circ\)) are shown as horizontal grey reference lines. The light grey shaded band marks the observed Crab torus extent \(\lambda \lesssim 30^\circ\) \cite{Komissarov2003, Hester1995, Hester2008, Weisskopf2000}.}
  \label{fig:latitude}
\end{figure}

\textit{Parameter-free prediction}:
Geometric modelling of the \textit{Chandra} X-ray torus places the Crab spin axis at approximately $\zeta \approx 61^\circ$ to the plane of the sky, implying an observed torus extent of $\lambda \sim 29^\circ$ \cite{NgRomani2004, NgRomani2008, Hester1995, Komissarov2003, Hester2008}. For an ultrarelativistic equation of state $\Gamma = 4/3$, Eq.~(\ref{eq:chiinf}) predicts $\chi_{\infty} = \arcsin \frac{2-\Gamma}{\Gamma} \sim 30^\circ$, in excellent agreement with the observed termination-shock geometry. Since this prediction depends only on the equation of state, it constitutes a parameter-free consequence of our theory.
\begin{figure}
  \centering
  \includegraphics[width=\columnwidth]{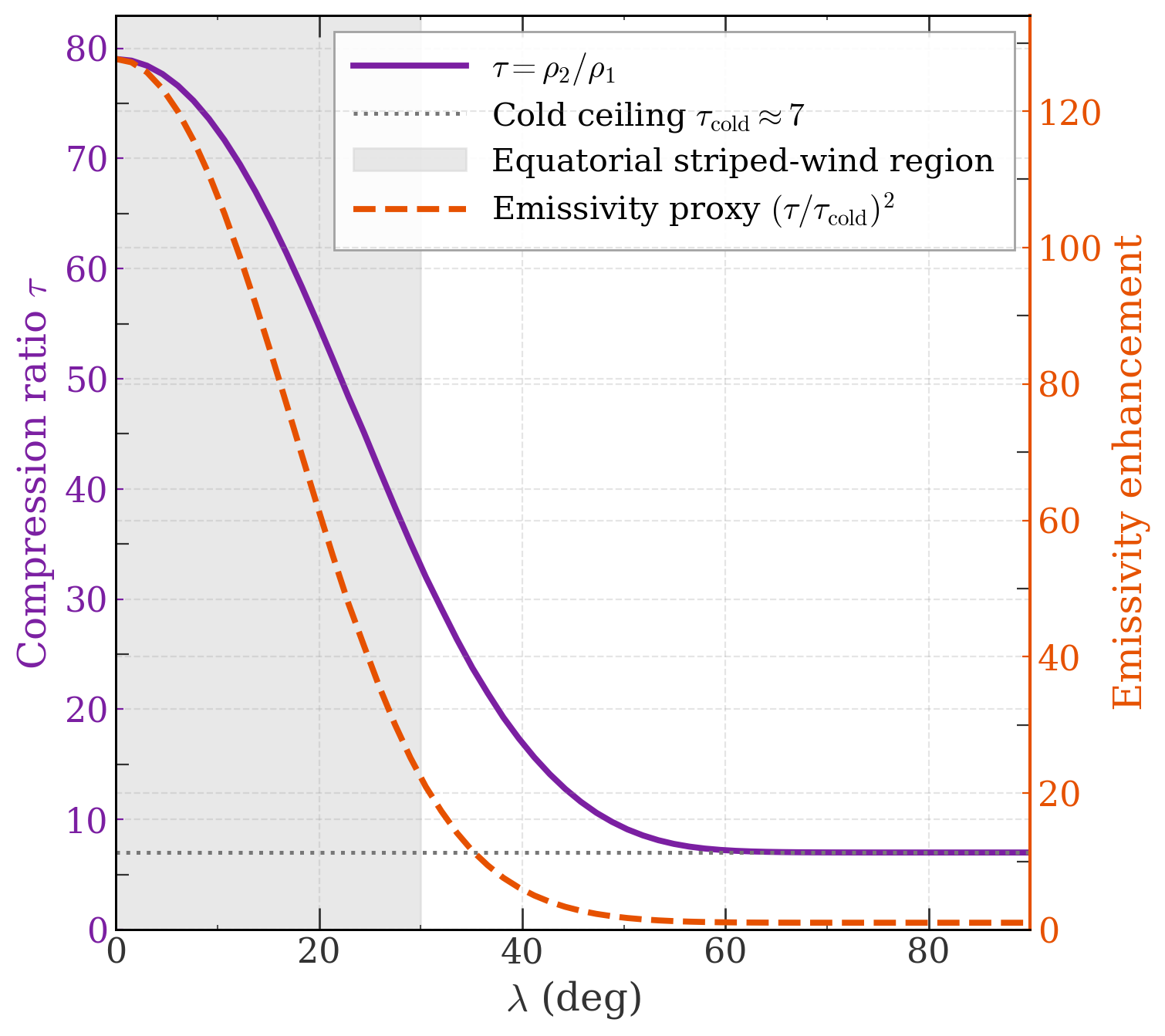}
  \caption{Equatorial density spike from thermal ceiling breakdown. Left axis (purple) shows the shock compression ratio \(\tau(\lambda)\) at the detachment point for \(\alpha_{\mathrm{eq}} = 0.1\). The cold Rankine-Hugoniot ceiling \(\tau_c \approx 7\) is marked as a grey dotted horizontal line. Right axis (orange dashed) shows the emissivity proxy \((\tau/\tau_c)^2\). The light grey band marks the equatorial striped wind zone with \(\lambda \lesssim 30^\circ\).}
  \label{fig:compression_crab}
\end{figure}

\textit{Equatorial thermal heating}:
The equatorial striped wind contains alternating magnetic polarity where reconnection converts magnetic energy into particle thermal energy \cite{Lyubarsky2003, Sironi2011, Cerutti2020, Lyubarsky2010, Petri2012}. Consequently, $\alpha_1>0$ is expected near the equator. We model the latitude dependence as
\begin{equation}
\alpha_1(\lambda)=\alpha_{1\mathrm{eq}}
\exp\left(-\frac{\lambda^2}{\sigma^2}\right),
\qquad
\sigma=20^\circ,
\label{eq:alpha_profile}
\end{equation}
with $\alpha_{1\mathrm{eq}}=0.1$ and $0.05$, representing strong and moderate heating, respectively. The adopted width is representative of the striped-wind reconnection zone \cite{Lyubarsky2003, Kirk2009}.
For illustration, we take $\alpha_{1\mathrm{eq}}=0.1$ and $M_1=100$, a conservative choice given estimates of the Crab wind Lorentz factor $\gamma_w\sim10^4$--$10^6$ \cite{Kennel1984, Kirk2009}, with the lower end expected for striped-wind models \cite{Lyubarsky2003, Kirk2009}. At the equatorial detachment point, the compression ratio exceeds the cold Rankine--Hugoniot limit $\tau_{\mathrm{cold}}\approx7$ by more than an order of magnitude. Since optically thin synchrotron emission satisfies $j_\nu\propto nB^{1+\alpha_s}\propto\tau^{2+\alpha_s}$ for a perpendicular shock ($B\propto\tau$), with $\alpha_s \simeq 1$, the emissivity is enhanced by at least $\mathcal{E} \gtrsim (\tau/\tau_{\mathrm{cold}})^2 \gtrsim 100$
where the quadratic scaling is a conservative lower bound. Such enhancements are inaccessible to cold hydrodynamic models within the present framework.
Figure~\ref{fig:latitude} shows that $\chim$ is systematically reduced throughout the observed Crab torus relative to the cold-fluid prediction, implying that cold models overestimate the shock detachment threshold. Figure~\ref{fig:compression_crab} shows the corresponding enhancement of the compression ratio and emissivity proxy $(\tau/\tau_{\mathrm{cold}})^2$ across the striped-wind region, naturally accounting for the bright inner torus within our thermal-shock framework \cite{Porth2014, Camus2009}. Future high-resolution \textit{Chandra} imaging of the torus boundary, together with spectral constraints on the upstream thermal parameter $\alpha_1$, can directly test this prediction.

\begin{figure}
  \centering
  \includegraphics[width=\columnwidth]{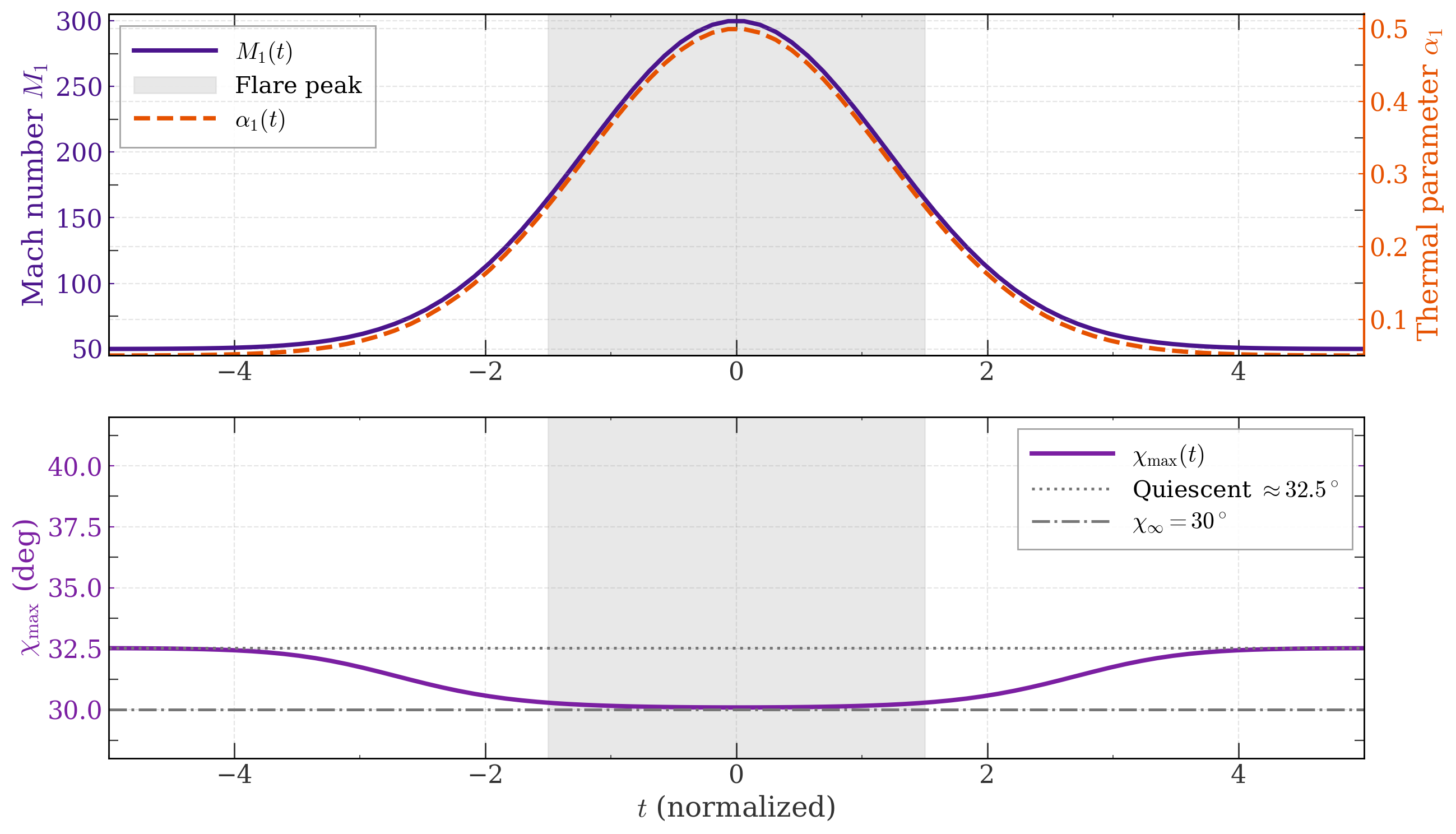}
  \caption{Transient flare evolution. Top panel shows the upstream Mach number 
  \(M_1(t)\) (purple solid) and the thermal parameter \(\alpha_1(t)\) (orange dashed) as 
  functions of normalized time. The light grey band indicates the main flare 
  phase. Bottom panel shows the instantaneous detachment angle \(\chi_{\text{max}}(t)\) 
  (purple solid), the quiescent state value at about \(32.5^\circ\) (horizontal grey dotted), 
  and the universal asymptote \(30^\circ\) (horizontal grey dash-dot).}
  \label{fig:flare_crab}
\end{figure}

\textit{$\gamma$-ray flares and time-dependent shock geometry}:
The physical origin of Crab $\gamma$-ray flares remains uncertain, although magnetic reconnection near the termination shock is widely regarded as the leading mechanism \cite{Cerutti2014, Lyutikov2018, Tavani2011, Abdo2011}. Reconnection is expected to convert magnetic energy into both bulk kinetic energy and thermal pressure, leading to a transient increase in the upstream Mach number $M_1$ and thermal parameter $\alpha_1$. To illustrate the resulting evolution of the shock geometry, we model both quantities by synchronized Gaussian profiles,
\begin{align}
M_1(t) &= M_{\rm base}+(M_{\rm peak}-M_{\rm base})
e^{-t^2/2\sigma_t^2}, \\
\alpha_1(t) &= \alpha_{1,\rm base}
+(\alpha_{1,\rm peak}-\alpha_{1,\rm base})
e^{-t^2/2\sigma_t^2},
\end{align}
using representative values $M_{\rm base}=50, M_{\rm peak}=300, \alpha_{1,\rm base}=0.05, \alpha_{1,\rm peak}=0.5$, and $\sigma_t=1.2$, corresponding to the observed several-day flare duration \cite{Tavani2011, Abdo2011}.
Figure~\ref{fig:flare_crab} shows that the increase in $M_1$, which alone tends to increase $\chi_{\infty}$, competes with the thermal increase in $\alpha_1$, which raises the turning parameter $R$ and suppresses the deflection angle. During the flare, the thermal effect dominates, producing a transient reduction in $\chi_{\infty}$. As the flare subsides, both $M_1$ and $\alpha_1$ return to their quiescent values and the detachment angle recovers smoothly without overshooting. At the flare peak ($M_1=300, \alpha_1=0.5$), the system approaches the ultra-thermal, ultra-relativistic regime, causing $\chim$ to approach, but not reach, the asymptotic value $\chinf$, consistent with the finite $M_1$ behaviour derived in Sec.~\ref{app:ultra}.
These results suggest that transient heating events should produce time-dependent shock morphology. In particular, simultaneous \textit{Chandra} and \textit{Fermi}-LAT observations during Crab flares could reveal a temporary narrowing of the X-ray torus together with enhanced emission from the inner ring, signatures absent in cold-wind models.

\textit{Summary}: 
Our framework yields three principal results for the Crab Nebula. First, the parameter-free prediction $\chi_{\infty}(4/3)$ agrees with the observed torus extent \cite{NgRomani2004, NgRomani2008, Hester1995, Komissarov2003, Hester2008, Weisskopf2000}. Second, finite upstream heating suppresses the detachment angle while enhancing the downstream compression, providing a thermodynamic explanation for the bright inner ring. Finally, observations of the termination-shock geometry offer a direct means of constraining the turning parameter $R$, thereby linking observable morphology to the thermodynamic state of relativistic shocks. A complete treatment requires relativistic MHD oblique shock theory with finite temperature and variable magnetization \cite{Begelman1992, Lyubarsky2003, Porth2014, Camus2009}, which we leave for future work.

\section{Conclusion}
\label{sec:conclusion}
In this work, we developed a thermodynamic framework for relativistic oblique shocks at finite temperature based on a single dimensionless quantity, the turning parameter, $R = (h_2/\tau)/h_1$, which encapsulates the equation of state, upstream Mach number, and thermal state of the flow. The entire shock geometry is governed by the compact turning relation, $\sin(2\phi-\chi) = [(1+R)/(1-R)]\sin\chi$
establishing $R$ as the fundamental variable controlling relativistic shock deflection.
Starting from the Taub adiabat, we derived a master implicit equation that exactly recovers the cold Rankine-Hugoniot compression ratio in the limit $\alpha_1 \to 0$. A first-order expansion in the upstream thermal parameter showed that any finite temperature suppresses the maximum deflection angle ($\chi_1 < 0$), implying that cold-fluid models systematically overestimate shock attachment. In the combined ultra-thermal and ultra-relativistic limit, the turning parameter saturates to $\Rinf = \Gamma - 1$, yielding the universal asymptotic detachment angle $\chinf = \arcsin[(2-\Gamma)/\Gamma]$,
which depends only on the equation of state. The same saturation naturally recovers the Shi et al. minimum-intensity locus, $\phi = \chi/2 + \pi/4$, thereby providing its first thermodynamic interpretation. 

Numerical shock-polar calculations confirm all analytical predictions and reveal a non-monotonic dependence of $\chi_{\max}$ on the Mach number in hot relativistic flows, a feature absent in the cold limit.
As an illustrative application, we applied the framework to the Crab pulsar wind nebula. For an ultrarelativistic equation of state ($\Gamma=4/3$), the parameter-free prediction $\chinf(4/3) = 30^\circ$ agrees with the observed torus extent. Finite upstream heating suppresses the detachment angle while enhancing the compression ratio, providing a simple thermodynamic interpretation of the bright equatorial ring. Since the Crab wind is magnetically dominated, these results should be regarded as illustrative rather than a complete physical description; extending the present formalism to relativistic magnetohydrodynamics remains an important direction for future work.
More generally, the turning parameter provides a direct connection between observable shock geometry and the underlying thermodynamic state of relativistic flows. Measurements of shock deflection can therefore constrain R, offering a simple diagnostic of the equation of state and thermal content in relativistic shocks. We anticipate that this framework will be useful for interpreting shocks in pulsar wind nebulae \cite{Gaensler2002, Buhler2014, DiPalma2017, Weisskopf2000, Hester2008, Dinsmore2024, Olmi2024, Barkov2019, Toropina2019, Posselt2017, Gagnon2026}, relativistic jets \cite{Senno2016, Senno2017, Zhang2021, Gutierrez2025, Peretti2026, SaizPerez2025}, supernova remnants \cite{Kobayakawa2002, Bykov2024, Bykov2025, Kamijima2024, Petruk2026, Cristofari2025}, and other high-energy astrophysical systems \cite{Decker2008, Cissoko1997, Richardson2008, Camus2009, Prete2026, Tang2025}, while also providing a natural starting point for future extensions that include magnetic fields \cite{Begelman1992, Lyubarsky2003, Lu2021, Guo2024, Morikawa2024}, anisotropic plasmas \cite{Lyubarsky2010}, and radiative processes \cite{Sironi2009, Sironi2011, Clarisse2026, Lemoine2025}.
\begin{table}[h]
\caption{Key analytic results for relativistic oblique-shock detachment.\label{tab:summary}}
\begin{ruledtabular}
\begin{tabular}{llc}
Quantity & Formula & Value ($\Gamma=5/3$) \\
Turning parameter $R$ & $(h_2/\tau)/h_1$ & --- \\
$\tau_0$ (strong-shock limit) & $(\Gamma+1)/(\Gamma-1)$ & $4$ \\
$R_0$ (strong-shock limit) & $(\Gamma-1)/(\Gamma+1)$ & $1/4$ \\
$\Rinf$ (ultra-thermal) & $\Gamma-1$ & $2/3$ \\
$\chinf$ (ultra-thermal) & $\arcsin[(2-\Gamma)/\Gamma]$ & $11.54^\circ$ \\
\end{tabular}
\end{ruledtabular}
\end{table}

\begin{acknowledgments}
RM acknowledges the support from the Indian Institute for Science Education and Research Bhopal for providing constant infrastructure and research support. RAS acknowledges the Indian Institute of Technology Indore for providing the academic environment during the course of this work. AV gratefully acknowledges the Indian Institute of Science Education and Research (IISER) Bhopal and the Indian Institute of Astrophysics (IIA), Bengaluru, for generously providing the necessary research infrastructure and facilities that made this work possible. 

\end{acknowledgments}

\section*{DATA AVAILABILITY}
This paper has no additional data as it is a theoretical
work.
\appendix
\section{Derivation of the Taub Adiabat}
\label{app:taub}
The conservation laws~\eqref{eq:baryon},~\eqref{eq:tangmom}, and~\eqref{eq:energy} define three kinematic invariants: $j\equiv\rho_i u^n_i$, $L\equiv h_i u^t_i$, and $K\equiv h_i\gamma_i$. Substituting these into the four-velocity normalization $\gamma_i^2 = 1+(u^n_i)^2+(u^t_i)^2$ yields
\begin{equation}
  K^2 = h_i^2 + \frac{j^2 h_i^2}{\rho_i^2} + L^2.
  \label{eq:A2}
\end{equation}
Evaluating~\eqref{eq:A2} for the upstream ($i=1$) and downstream ($i=2$) states and taking their difference eliminates $K^2$ and $L^2$:
\begin{equation}
  j^2\!\left(\frac{h_1^2}{\rho_1^2} - \frac{h_2^2}{\rho_2^2}\right) = h_2^2 - h_1^2.
  \label{eq:A4}
\end{equation}
Factoring the left-hand side as a difference of squares and substituting the normal-momentum condition~\eqref{eq:normalmom}, $j^2(h_1/\rho_1 - h_2/\rho_2)=p_2-p_1$, gives
\begin{equation}
  (p_2-p_1)\!\left(\frac{h_1}{\rho_1}+\frac{h_2}{\rho_2}\right) = h_2^2-h_1^2.
  \label{eq:A6}
\end{equation}
Substituting the compression ratio $\tau=\rho_2/\rho_1$ into the left side and dividing by $\rho_1$ yields the final Taub adiabat:
\begin{equation}
  h_2^2-h_1^2 = \frac{p_2-p_1}{\rho_1}\!\left(h_1+\frac{h_2}{\tau}\right).
  \label{eq:A7}
\end{equation}
\section{The Master Implicit Equation}
\label{app:master}
Inserting the specific enthalpies $h_1=1+a\alpha_1$, $h_2=1+a\xi\alpha_1/\tau$, and the pressure difference $(p_2-p_1)/\rho_1=\alpha_1(\xi-1)$ directly into the Taub adiabat~\eqref{eq:A7} yields the master implicit equation:
\begin{equation}
\begin{split}
\left(1+\frac{a\xi\alpha_1}{\tau}\right)^{\!2}-(1+a\alpha_1)^2
&= \alpha_1(\xi-1)\Bigl[(1+a\alpha_1) \\
&\quad + \frac{1}{\tau}\!\left(1+\frac{a\xi\alpha_1}{\tau}\right)\Bigr],
\end{split}
\label{eq:B6}
\end{equation}
The pressure ratio $\xi$ is obtained self-consistently from the normal-momentum condition~\eqref{eq:normalmom}. Using $j^2=\rho_1^2(u^n_1)^2=\rho_1^2\Mn^2/h_1^2$ (where $u^n_1=\Mn/h_1$ in the shock rest frame), substituting into~\eqref{eq:normalmom} alongside $\rho_2=\tau\rho_1$ and $p_i=\rho_i h_i\alpha_1/a$, and rearranging yields:
\begin{equation}
  \xi = \frac{\Gamma\Mn^2\bigl(1-1/(\tau h_1)\bigr)+1}
             {1+\Gamma\Mn^2 a\alpha_1/(\tau^2 h_1)}.
  \label{eq:B7}
\end{equation}
Equations~\eqref{eq:B6} and~\eqref{eq:B7} are solved iteratively starting from cold-limit values until convergence.
\section{Derivation of the Turning Relation}
\label{app:turning}
With $\phi$ measured from the shock surface (see Fig.~\ref{fig:geometry}), the upstream and downstream velocity decompositions are:
\begin{align}
  u^n_1 &= \gamma_1 v_1\sin\phi, & u^t_1 &= \gamma_1 v_1\cos\phi, \label{eq:C1}\\
  u^n_2 &= \gamma_2 v_2\sin(\phi-\chi), & u^t_2 &= \gamma_2 v_2\cos(\phi-\chi). \label{eq:C2}
\end{align}
Substituting these into tangential-momentum continuity~\eqref{eq:tangmom} and baryon conservation~\eqref{eq:baryon}, and dividing the former by the latter cancels the $\gamma_i v_i$ factors to yield:
\begin{equation}
  h_1\cot\phi = \frac{h_2}{\tau}\cot(\phi-\chi).
  \label{eq:C6}
\end{equation}
Introducing the turning parameter $R\equiv(h_2/\tau)/h_1$, dividing~\eqref{eq:C6} by $h_1$, and cross-multiplying gives:
\begin{equation}
  \cos\phi\sin(\phi-\chi) - R\sin\phi\cos(\phi-\chi) = 0.
  \label{eq:C7}
\end{equation}
Applying the standard product-to-sum trigonometric identities ($\cos A\sin B$ and $\sin A\cos B$) to both terms and collecting $\sin(2\phi-\chi)$ on the left and $\sin\chi$ on the right immediately yields the turning relation:
\begin{equation}
  \sin(2\phi-\chi) = \frac{1+R}{1-R}\sin\chi.
  \label{eq:C12}
\end{equation}
Setting $\alpha_1=0$ gives $h_1=h_2=1$, hence $R=1/\tau_0$. The prefactor then becomes $(\tau_0+1)/(\tau_0-1)$, yielding the cold turning relation:
\begin{equation}
  \sin(2\phi-\chi) = \frac{\tau_0+1}{\tau_0-1}\sin\chi.
  \label{eq:turning_cold}
\end{equation}
\section{The Detachment Stationarity Condition}
\label{app:detach}
Define $F(\phi,\chi)\equiv\sin(2\phi-\chi)-[(1+R)/(1-R)]\sin\chi$. Differentiating with respect to $\phi$ at fixed $\chi$ requires applying the chain rule to the prefactor:
\begin{equation}
  \frac{\partial}{\partial\phi}\!\left(\frac{1+R}{1-R}\right) = \frac{2}{(1-R)^2}\frac{\partial R}{\partial\phi}.
  \label{eq:D2}
\end{equation}
Substituting this derivative into $\partial F/\partial\phi = 0$ immediately yields the stationarity condition:
\begin{equation}
  2\cos(2\phi-\chi) = \frac{2}{(1-R)^2}\frac{\partial R}{\partial\phi}\sin\chi.
  \label{eq:D3}
\end{equation}
\section{Cold Limit: Rankine-Hugoniot Compression Ratio}
\label{app:cold}
Setting $\alpha_1\to0$ implies $h_1\to1$, $h_2\to1$, and $\tau\to\tau_0$. Expanding the master equation~\eqref{eq:B6} to $O(\alpha_1)$ and matching coefficients yields:
\begin{equation}
  2a\!\left(\frac{\xi}{\tau_0}-1\right) = (\xi-1)\!\left(1+\frac{1}{\tau_0}\right).
  \label{eq:E3}
\end{equation}
Solving for $1/\tau_0$ gives:
\begin{equation}
  \frac{1}{\tau_0} = \frac{(\xi-1)+2a}{2a\xi-(\xi-1)}.
  \label{eq:E4}
\end{equation}
Substituting the cold-limit pressure ratio $\xi_{\rm cold}=[2\Gamma\Mn^2-(\Gamma-1)]/(\Gamma+1)$ and the EOS constant $a=\Gamma/(\Gamma-1)$ into~\eqref{eq:E4} simplifies directly to the classical Rankine-Hugoniot compression ratio:
\begin{equation}
  \tau_0 = \frac{(\Gamma+1)\Mn^2}{(\Gamma-1)\Mn^2+2}.
  \label{eq:E8}
\end{equation}
In the strong-shock limit ($\Mn\to\infty$), this bounds the compression to $\tau_0\to(\Gamma+1)/(\Gamma-1)$, which equals $4$ for $\Gamma=5/3$ and $7$ for $\Gamma=4/3$.
\section{First-Order Perturbative Expansion}
\label{app:perturb}
Write $\tau=\tau_0+\alpha_1\tau_1+O(\alpha_1^2)$, $R=R_0+\alpha_1 R_1+O(\alpha_1^2)$, and $\chim=\chi_0+\alpha_1\chi_1+O(\alpha_1^2)$, with $R_0=1/\tau_0$.
\subsection{Compression Correction $\tau_1$}
Expanding the terms in the master equation~\eqref{eq:B6} to $O(\alpha_1^2)$ and isolating the $O(\alpha_1^2)$ coefficients requires expanding $h_2^2-h_1^2$ against the right-hand side. Grouping the $\tau_1$ terms and simplifying yields the compression correction:
\begin{equation}
  \tau_1 = \frac{a\tau_0^2(\xi-1+a)+a\xi[(\xi-1)-a\xi]}{(\xi-1)-2a\xi},
  \label{eq:F13}
\end{equation}
where $\xi=\xi_{\rm cold}$ at zeroth order.
\subsection{Turning-Parameter Correction $R_1$}
Expanding $R = (h_2/\tau)/h_1$ to first order using the expansion of $1/\tau$ and $h_1^{-1}=1-a\alpha_1+O(\alpha_1^2)$ gives:
\begin{equation}
  R_1 = \frac{a\xi-\tau_1}{\tau_0^2} - \frac{a}{\tau_0}.
  \label{eq:F17}
\end{equation}
\subsection{Detachment-Angle Correction $\chi_1$}
Define $G(\phi,\chi,R)\equiv\sin(2\phi-\chi)-[(1+R)/(1-R)]\sin\chi$. At the cold detachment point $(\phi_0,\chi_0,R_0)$, $G=0$ and $\partial G/\partial\phi=0$. Setting $\phi=\phi_0+\alpha_1\phi_1$, $\chi=\chi_0+\alpha_1\chi_1$, and $R=R_0+\alpha_1 R_1$, a Taylor expansion to $O(\alpha_1)$ gives:
\begin{equation}
  \frac{\partial G}{\partial\chi}\bigg|_0\chi_1
  +\frac{\partial G}{\partial R}\bigg|_0 R_1 = 0,
  \label{eq:F19}
\end{equation}
since the $\phi_1$ term vanishes via the stationarity condition. Evaluating the partial derivatives and solving for $\chi_1$ yields:
\begin{equation}
  \chi_1 = -K\,R_1,
  \label{eq:F22}
\end{equation}
where:
\begin{equation}
  K \equiv \frac{2\sin\chi_0/(1-R_0)^2}
                 {\cos(2\phi_0-\chi_0)+\dfrac{1+R_0}{1-R_0}\cos\chi_0} > 0.
  \label{eq:F23}
\end{equation}
Since $K>0$ and $R_1>0$ for all physical strong shocks, $\chi_1<0$.
\section{Ultra-Thermal Limit}
\label{app:ultra}
\subsection{Saturated Turning Parameter $\Rinf$}
For $\alpha_1\to\infty$, rest-mass terms become negligible ($h_1\approx a\alpha_1$, $h_2\approx a\xi\alpha_1/\tau$). The $\alpha_1$ factors cancel in the turning parameter $R$:
\begin{equation}
  R \approx \frac{\xi}{\tau^2} \equiv Y.
  \label{eq:G2}
\end{equation}
Substituting these approximations into the Taub adiabat~\eqref{eq:A7} and collecting terms in $\xi$ yields:
\begin{equation}
  \xi(aY-1-Y) = a-1-Y.
  \label{eq:G6}
\end{equation}
Taking the limit $\Mn\to\infty$ forces the pressure ratio $\xi\to\infty$. For~\eqref{eq:G6} to remain finite, the coefficient of $\xi$ must vanish, yielding $Y=1/(a-1)$. Since $a=\Gamma/(\Gamma-1)$, we obtain the saturated turning parameter:
\begin{equation}
  \Rinf = \Gamma-1.
  \label{eq:G9}
\end{equation}
For $\Gamma=5/3$, $\Rinf=2/3$; for $\Gamma=4/3$, $\Rinf=1/3$.
\subsection{Universal Detachment Angle $\chinf$}
With $R=\Gamma-1=\mathrm{const}$, the derivative $\partial R/\partial\phi$ vanishes. The stationarity condition~\eqref{eq:D3} reduces purely to geometry ($\cos(2\phi-\chi) = 0$), yielding the minimum-intensity locus:
\begin{equation}
  \phi = \frac{\chi}{2}+\frac{\pi}{4}.
  \label{eq:G11}
\end{equation}
Substituting this locus and $R=\Gamma-1$ into the turning relation~\eqref{eq:C12} gives $\sin(\pi/2) = [\Gamma/(2-\Gamma)]\sin\chinf$. Solving for $\chinf$ gives the universal detachment angle:
\begin{equation}
  \chinf = \arcsin\!\left(\frac{2-\Gamma}{\Gamma}\right).
  \label{eq:G14}
\end{equation}
For $\Gamma=5/3$, $\chinf=11.54^\circ$; for $\Gamma=4/3$, $\chinf=30.00^\circ$.
\section{Equivalence with the Landau-Lifshitz Mach-Number Relation}
\label{app:ll}
Using the cold compression ratio from Eq.~\eqref{eq:E8}, we can write the prefactor as:
\begin{equation}
  \frac{\tau_0+1}{\tau_0-1} = \frac{\Gamma\Mn^2+1}{\Mn^2-1}.
  \label{eq:H1}
\end{equation}
Substituting this into the cold turning relation~\eqref{eq:turning_cold} and cross-multiplying to isolate the Mach number terms gives:


\begin{equation}
\Mn^2\bigl[\sin(2\phi-\chi)-\Gamma\sin\chi\bigr] = \sin(2\phi-\chi)+\sin\chi.
\end{equation}

Solving directly for $\Mn^2$ recovers Eq.~(\ref{eq:ll_mach}), the standard oblique-shock Mach-number relation~\cite{Landau1987, Shi2020}:
\begin{equation}
  \Mn^2 = 1 + \frac{(\Gamma+1)\sin\chi}{\sin(2\phi-\chi)-\Gamma\sin\chi}.
\end{equation}

\section{Supplementary Shock-Polar and Compression Figures}
\label{app:extra_figs}
\begin{center}
Supplementary figures for Sec.~\ref{sec:numerics}.
\end{center}

\begin{figure}[H]
  \centering
  \includegraphics[width=0.85\columnwidth]{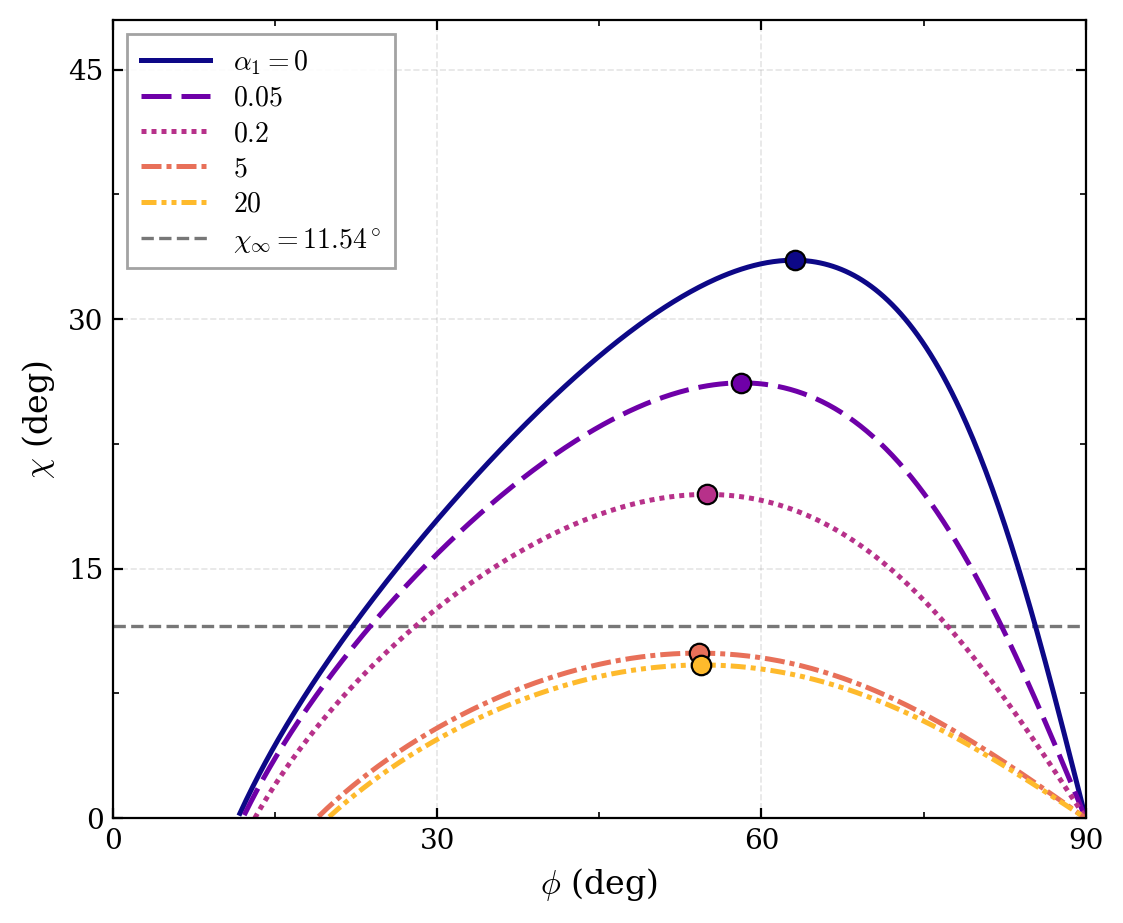}
  \caption{Shock polar ($\chi$ vs.\ $\phi$) for $M_1=5$, $\Gamma=5/3$, and
    five upstream thermal parameters. Filled circles mark $\chim$; the horizontal dashed
    line is $\chinf=11.54^\circ$.  The progressive shrinkage of successive
    shock polars directly visualizes thermal suppression through the turning
    parameter $R$.  Hot shock polars ($\alpha_1=5$, 20) peak below $\chinf$ because
    the combined ultra-thermal limit is not reached at finite $M_1=5$. This single-Mach-number
    slice through Fig.~\ref{fig:polargrid} isolates the thermal-suppression trend discussed
    in Sec.~\ref{sec:numerics}.}
  \label{fig:polar}
\end{figure}

\begin{figure}[h]
  \centering
  \includegraphics[width=0.85\columnwidth]{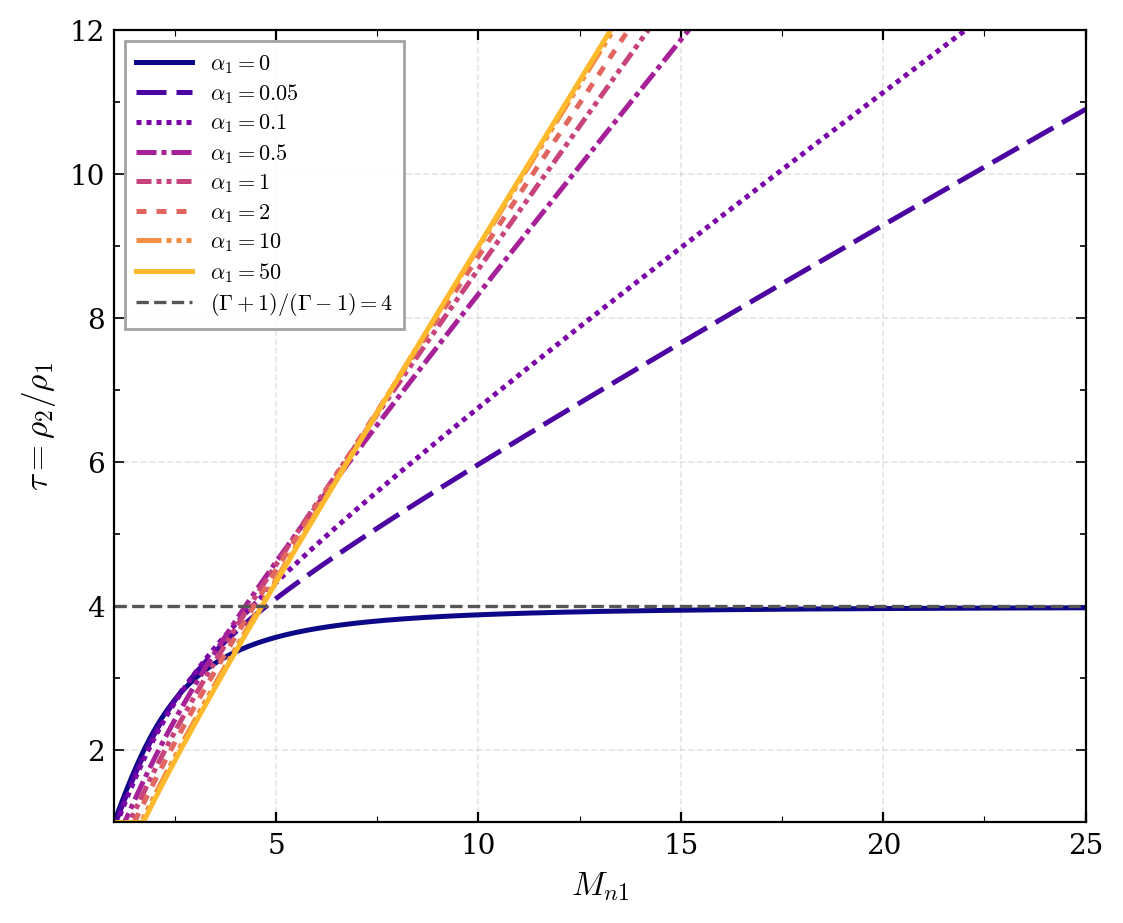}
  \caption{Compression ratio $\tau=\rho_2/\rho_1$ vs.\ $\Mn$ for eight
    values of $\alpha_1$ ($\Gamma=5/3$).  The cold saturation ceiling at
    $(\Gamma+1)/(\Gamma-1)=4$ (horizontal dashed) is progressively lifted; hotter gas
    sustains unbounded growth $\tau\propto\Mn$. The gradual bending of each curve away
    from the cold ceiling marks the Mach number above which thermal enthalpy generation,
    rather than compressional heating, dominates the downstream state.}
  \label{fig:compression}
\end{figure}

\bibliographystyle{apsrev4-2}
\bibliography{references}
\end{document}